# Relationship between charge redistribution and ferromagnetism at the heterointerface between perovskite oxides LaNiO$_3$ and LaMnO$_3$

Miho Kitamura,[1,2] Masaki Kobayashi,[1] Enju Sakai,[1] Makoto Minohara,[1] Ryu Yukawa,[1] Daisuke Shiga,[1] Kenta Amemiya,[1] Yosuke Nonaka,[3] Goro Shibata,[3] Atsushi Fujimori,[3] Hiroshi Fujioka,[2] Koji Horiba,[1,a)] and Hiroshi Kumigashira[1]

[1]*Photon Factory, Institute of Materials Structure Science, High Energy Accelerator Research Organization (KEK), 1-1 Oho, Tsukuba 305-0801, Japan*

[2]*Institute of Industrial Science, The University of Tokyo, 4-6-1 Komaba, Meguro-ku, Tokyo 153-8505, Japan*

[3]*Department of Physics, Graduate School of Science, The University of Tokyo, 7-3-1 Hongo, Bunkyo-ku, Tokyo 113-0033, Japan*

[a)]Author to whom correspondence should be addressed.　Email: horiba@post.kek.jp

*Abstract*

To investigate the relationship between the charge redistribution and ferromagnetism at the heterointerface between perovskite transition-metal oxides $LaNiO_3$ (LNO) and $LaMnO_3$ (LMO), we performed x-ray absorption spectroscopy and x-ray magnetic circular dichroism (XMCD) measurements. In the LNO/LMO heterostructures with asymmetric charge redistribution, the electrons donated from Mn to Ni ions are confined within one monolayer (ML) of LNO at the interface, whereas holes are distributed over 3–4 ML on the LMO side. A detailed analysis of the Ni-$L_{2,3}$ and Mn-$L_{2,3}$ XMCD spectra reveals that Ni magnetization is induced only by the $Ni^{2+}$ ions in the 1 ML LNO adjacent to the interface, while the magnetization of Mn ions is increased in the 3–4 ML LMO of the interfacial region. The characteristic length scale of the emergent (increased) interfacial ferromagnetism of the LNO (LMO) layers is in good agreement with that of the charge distribution across the interface, indicating a close relationship between the charge redistribution due to the interfacial charge transfer and the ferromagnetism of the LNO/LMO interface. Furthermore, the XMCD spectra clearly demonstrate that the vectors of induced magnetization of both ions are aligned ferromagnetically, suggesting that the delicate balance between the exchange interactions occurring inside each layer and across the interface may induce the canted ferromagnetism of $Ni^{2+}$ ions, resulting in weak magnetization in the 1 ML LNO adjacent to the interface.

**I. INTRODUCTION**

Heterostructures composed of different perovskite transition-metal oxides have attracted considerable attention because of the wide variety of novel electronic and magnetic properties that emerge at their interfaces, which cannot be accomplished in the bulk constituents [1, 2]. One of the most interesting properties is the ferromagnetism at the heterointerface formed between "non-magnetic" compounds [3]. Some oxide superlattices composed of antiferromagnetic insulators and/or paramagnetic metals exhibit ferromagnetic properties, such as $LaMnO_3$ (LMO)/$SrMnO_3$ [4-6], $CaMnO_3$/$CaRuO_3$ [7-9], $CaMnO_3$/$LaNiO_3$ (LNO) [10, 11], $LaFeO_3$/$LaCrO_3$ [12, 13], and $LaFeO_3$/LMO [14].

The charge transfer across heterointerfaces is one of the most important factors responsible for this exotic interfacial ferromagnetism [3]. The valence changes of the constituent transition-metal ions stemming from the interfacial charge transfer, accompanied by the reconstructions of the orbital and spin degrees of freedom, have a significant effect on magnetic exchange interactions. In addition, to better understand the novel magnetic properties emerging at the heterointerfaces, it is important to elucidate not only the valence change but also the spatial distribution of the transferred charges. For example, for LMO/$SrMnO_3$ superlattices, the balance between the length of the charge distribution and the period of superlattices determines the properties of the resulting complex magnetic structures. Spatially uniform magnetic properties, such as those of $La_{1-x}Sr_xMnO_3$ alloys, emerge for superlattices with short periods, whereas phase-separated magnetic structures emerge for those with longer periods [4]. Calculations based on the dynamical mean-field theory have also predicted that different magnetic phases form depending on the charge distribution length [15].

Concerning the LNO/LMO heterointerface, the valence changes in transition-metal ions due to the charge transfer should significantly affect the interfacial ferromagnetism [16-18].

The charge transfer $Ni^{3+} + Mn^{3+} \rightarrow Ni^{2+} + Mn^{4+}$ has been reported to occur across the LNO/LMO heterointerface [17-20], analogously to the double perovskite $La_2NiMnO_6$ (LNMO) system [21-23]. As a result of this interfacial charge transfer, ferromagnetically coupled magnetization is induced between Ni and Mn ions [17-20], although the constituent LNO and LMO are a paramagnetic metal and an antiferromagnetic insulator in their bulk forms, respectively.

Recently, we revealed the existence of a characteristic spatial distribution in the transferred charges at the (001)-oriented LNO/LMO heterointerface [24]. A detailed analysis of the thickness dependence of x-ray absorption spectroscopy (XAS) results revealed that the spatial distributions of the transferred charges in the two layers are significantly different; specifically, the transferred electrons are confined inside the one monolayer (ML) LNO, while the holes are distributed over 3–4 ML LMO. Therefore, the origin of the interfacial ferromagnetism observed in the (001)-oriented LNO/LMO heterostructures should be addressed considering the spatial charge distribution in the charge transfer region.

In this study, we investigate the relationship between the charge redistribution due to the interfacial charge transfer and the resultant ferromagnetism at the LNO/LMO heterointerface by using XAS and x-ray magnetic circular dichroism (XMCD) measurements. The interface (surface) sensitivity and elemental selectivity of these spectroscopic techniques enable extracting the electronic and magnetic properties of each transition-metal ion in the vicinity of the interface. Systematic thickness-dependent XMCD measurements reveal that the spatial redistributions of the transferred charges greatly influence the emergence of ferromagnetism at the LNO/LMO heterointerface. Weak magnetization is observed only for the $Ni^{2+}$ ions in the 1 ML LNO adjacent to the heterointerface, owing to the confinement of transferred electrons. Meanwhile, the ferromagnetism of the LMO layer is higher in the 3–4

ML interfacial region than that in the inner region. The characteristic length scale of the emerging (increased) interfacial ferromagnetism for LNO (LMO) layers is in good agreement with the charge distribution across the interface, indicating a close relationship between the charge redistribution due to the interfacial charge transfer and the ferromagnetism of LNO/LMO heterointerfaces. The magnetic moments of both ions couple ferromagnetically through superexchange (SE) interactions, thus weakly magnetizing the $Ni^{2+}$ ions, probably due to spin canting in the 1 ML LNO at the interface. These results suggest that the delicate balance between the exchange interactions inside each layer and across the interface may lead to the complex magnetic structures at the interface of transition-metal oxides.

## II. EXPERIMENT

Digitally controlled LMO/LNO/LMO and LNO/LMO/LNO trilayer structures were fabricated on atomically flat $TiO_2$-terminated $Nb:SrTiO_3$ (STO) (001) substrates inside the laser molecular-beam epitaxy chamber installed at beamline BL-2A MUSASHI of the Photon Factory, KEK (KEK-PF). The growth conditions are described in detail elsewhere [25, 26]. During LNO and LMO deposition, the substrate was maintained at temperatures of 450–500°C and 600–700°C, respectively, and the oxygen pressure was maintained at $1 \times 10^{-3}$ Torr. During growth, the thicknesses of the deposited LNO and LMO layers were precisely controlled by monitoring the intensity oscillation of the specular spot using reflection high-energy electron diffraction measurements. For the LMO/LNO/LMO trilayers, the thickness $n$ of the sandwiched LNO layer was varied from 2 to 5 ML, while the top and bottom LMO layer thicknesses were fixed at 5 and 20 ML, respectively. For the LNO/LMO/LNO trilayers, the thickness $m$ of the sandwiched LMO layer was varied between 2, 6, and 12 ML, while the thicknesses of the top and bottom LNO layers were fixed at 5 and 20 ML, respectively. The trilayer structures were subsequently annealed at 400°C for 45 min in oxygen at atmospheric pressure to allow oxygen to fill residual oxygen vacancies. The surface morphology of the prepared films was analyzed using *ex situ* atomic force microscopy in air. Atomically flat surfaces with step-and-terrace structures, which reflected the morphology of the Nb:STO substrate, were clearly observed for all samples, indicating that not only the surface but also the buried interfaces were atomically flat. All trilayer structures were fabricated under the same conditions as those of the previously reported bilayer structures, wherein chemically abrupt interfaces formed, with possible intermixing regions less than 1 ML thick, irrespective of their stacking order, as confirmed by high-angle annular dark-field scanning transmission electron microscopy observations [24].

XAS and XMCD measurements were carried out at BL-2A MUSASHI and BL-16A at KEK-PF, respectively. XAS spectra were taken at 300 K using horizontally polarized light, while XMCD spectra were obtained at 70 K using circularly polarized light under a magnetic field of 1 T. All XAS and XMCD spectra were obtained by measuring the sample drain current. During XMCD measurements, the angle between the incident x-ray beam and the sample normal was set to 60° to measure its in-plane magnetization. The polarization of the incident beam was kept constant, while the magnetic field was switched between positive and negative. The XMCD spectrum was defined as the difference between the spectra measured under positive and negative magnetic fields. The Ni-$L_{2,3}$ XAS and XMCD spectra were extracted by subtracting the contribution of the La-$M_4$ edge from the raw spectra because the Ni-$L_3$ edge partially overlapped with the very strong La-$M_4$ absorption edge owing to the close proximity of these two energy levels [24, 27].

**III. RESULTS & DISCUSSION**

Figure 1 shows the Ni-$L_{2,3}$ (Mn-$L_{2,3}$) XAS spectra of the sandwiched LNO (LMO) layer in the LMO/LNO/LMO (LNO/LMO/LNO) trilayer structures measured at 300 K for various thicknesses of each sandwiched layer. The spectra measured for the 20-ML LNO (LMO) film and an LNMO film [23, 28] are shown as references for the Ni$^{3+}$ (Mn$^{3+}$) and Ni$^{2+}$ (Mn$^{4+}$) states, respectively. Comparing the spectra of the sandwiched layers with the reference spectra clearly shows that the formal valence of Ni ions changes from 3+ in the LNO film to almost 2+ in the LMO/LNO (2 ML)/LMO trilayer film. In the counterpart LNO/LMO (2 ML)/LNO trilayer structure, the formal valence of Mn ions changes from 3+ in the LMO film to nearly 4+. These results clearly demonstrate that electrons transfer from the Mn ions to the Ni ions across the heterointerface, as reported in previous studies [17, 18, 20, 24]. As shown in Fig. 1(a), the spectra of the trilayer structures exhibit systematic changes with increasing $n$: The intensity of the characteristic sharp peak of the Ni-$L_3$ edge at about 853.5 eV corresponding to Ni$^{2+}$ species (originated from the interfacial charge transfer) gradually decreases. Simultaneously, the doublet of the Ni-$L_2$ edge centered at around 871.5 eV changes into a single peak, indicating that the relative fraction of Ni$^{3+}$ states increases with increasing $n$. By assuming that the transferred electrons are confined only to the 1 ML LNO adjacent to the interface [24] and the observed dependence of the XAS spectra on the thickness is well explained by a linear combination of Ni$^{2+}$ (LNMO) and Ni$^{3+}$ (LNO) states, it is suggested that the Ni$^{3+}$ states remain unchanged in the inner region of sandwiched LNO layers with $n > 2$.

As for the corresponding LNO/LMO/LNO trilayer structures [see Fig. 1(b)], we intentionally set the thicknesses of the sandwiched LMO layers to 2, 6, and 12 ML, assuming that holes are distributed within the 3–4 ML LMO adjacent to the interfaces [24]. When $m = 2$, both LMO layers are adjacent to the interfaces with the LNO layers, and when $m = 6$, all regions of the sandwiched LMO layer are expected to be affected by the charge transfer. In contrast, when $m = 12$, the LMO layer is expected to contain regions both affected and unaffected by the charge transfer. As shown in Fig. 1(b), the XAS spectrum measured for $m = 2$ is almost identical to that of LNMO ($Mn^{4+}$), suggesting that transferred holes are accommodated inside the sandwiched 2-ML LMO layer. When $m = 6$, the shoulder at 641.2 eV, indicating the existence of $Mn^{4+}$ states, becomes buried inside the main Mn-$L_3$ XAS peak centered at around 643.5 eV, and simultaneously, the Mn-$L_{2,3}$ XAS peaks become broader. These changes indicate that the average formal valence of Mn ions in the $m = 6$ trilayer structure decreases from 4+ of $m = 2$ due to hole distribution. After further increasing $m$ from 6 to 12 ML, the average formal valence of Mn ions becomes closer to 3+. This finding indicates the evolution of the $Mn^{3+}$ states in the inner region of the sandwiched LMO layer, which does not contribute to the charge transfer, as expected. These results are consistent with the asymmetric charge redistribution described in our previous report [24], wherein the electrons transferred from Mn ions to Ni ions are confined within the 1 ML LNO at the interface, whereas holes are distributed over 3–4 ML adjacent to the interface in the LMO layer.

Next, in order to investigate the relationship between the interfacial charge transfer and the emergence of ferromagnetism, we measured the circular dichroism of the XAS spectra for these trilayer structures. Figure 2(a) shows the Ni-$L_{2,3}$ XMCD spectra of the sandwiched LNO layers with various thicknesses $n$ in the LMO/LNO/LMO trilayers. The spectrum measured for the paramagnetic 20-ML LNO film is also shown as a reference. As expected, no XMCD signal is observed for the LNO film, indicating its paramagnetic properties [17]. In contrast, distinct XMCD signals are obtained for the Ni-$L_{2,3}$ edges of the LMO/LNO/LMO trilayers, whose intensities monotonically decrease with increasing $n$. These results indicate that the magnetization of Ni ions emerges at the heterointerface. As for the top and bottom

LMO layers of these trilayers, Mn-$L_{2,3}$ XMCD signals are clearly detected, as shown in the inset of Fig. 2(a). Notably, the signs of the Ni-$L_{2,3}$ and Mn-$L_{2,3}$ XMCD signals are the same, indicating that the magnetization vectors induced for both ions align ferromagnetically. This observed ferromagnetic coupling between Ni and Mn ions at the interface is consistent with the results of previous reports [17, 18, 29]. However, the length scale of the ferromagnetic region for interfacial LNO and LMO layers has not been determined yet due to experimental difficulties.

In order to overcome the challenges encountered in previous studies, we performed the line shape analysis of the Ni-$L_2$ XMCD spectra in Fig. 2(a). Here, the Ni-$L_2$ edges are selected for the analysis because the Ni-$L_2$ absorption edge does not overlap with the La-$M_4$ edge. Figure 2(b) shows the Ni-$L_2$ XMCD spectra of the LMO/LNO/LMO trilayers normalized by the maximum intensities to compare their spectral shapes. The line shapes of the Ni-$L_2$ XMCD spectra for all LMO/LNO/LMO trilayers remain almost unchanged within experimental accuracy. This invariance indicates that the electronic states of Ni ions contributing to the magnetization are the same, regardless of the thicknesses of the sandwiched LNO layers.

The obtained Ni-$L_2$ edge XMCD spectra closely resemble that of La$_2$Ni$^{2+}$Mn$^{4+}$O$_6$ [21], indicating that the magnetization emerging in the LNO layers is localized exclusively in Ni$^{2+}$ ions, which is consistent with the results of a previous study on LNO/LMO superlattices [17]. Considering that Ni$^{2+}$ states are present only in the 1 ML LNO at the interface [24], it can be concluded that magnetization is induced only in the Ni$^{2+}$ ions of the 1 ML LNO adjacent to the heterointerface as a result of the interfacial charge transfer.

To better understand the origins of the ferromagnetism emerging in the LNO layer, the spin moment induced in Ni$^{2+}$ ions should be investigated as a function of $n$. To quantitatively evaluate the spin moment of Ni$^{2+}$ ions, we performed an element-selective sum rule analysis [30,31] of the XMCD signal [17]. The effective spin moments averaged over Ni ions in the LMO/LNO/LMO trilayers are plotted in Fig. 2(c) for different values of $n$. The effective spin moment per Ni ion monotonically decreases with increasing $n$, indicating that magnetization emerges only around the heterointerface of the LMO layers, as described above. We simulated the averaged spin moment assuming that i) $Ni^{2+}$ ions exist only in the 1 ML LNO adjacent to the LMO interface, ii) $Ni^{2+}$ ions possess the full moment (2 $\mu_B$/Ni ion), which is expected for the $Ni^{2+}$ electron configuration ($3d^8$: $t_{2g}^6 e_g^2$), and iii) the $Ni^{3+}$ ions do not magnetize in the inner layers, away from the interface. The results of these simulations are also plotted in Fig. 2(c). Notably, the observed spin moments of interfacial $Ni^{2+}$ ions are significantly smaller than the full moments of the $Ni^{2+}$ states (2 $\mu_B$/Ni ion) and LNMO (1.9 $\mu_B$/Ni ion) [22]. However, in the present study, the induced moments are about 0.2–0.3 $\mu_B$/Ni ion, which is comparable with reported values, such as 0.12 $\mu_B$/Ni ion for an LNO (4 ML)/LMO (4 ML) superlattice [17] and 0.35 $\mu_B$/Ni ion for an LNO (4 ML)/LMO (2 ML) superlattice [18]. According to the Kanamori–Goodenough (KG) rules [32], magnetic moments should be absent from the LNO layer containing $Ni^{2+}$ ions because of the strong antiferromagnetic interactions between $Ni^{2+}$ ions, as discussed in detail below. These results therefore suggest that some ferromagnetic exchange interactions between Ni and Mn ions across the interface induce small but measurable spin moments in $Ni^{2+}$ ions.

To examine the behavior of the corresponding Mn ions, the Mn-$L_{2,3}$ XMCD spectra of the LNO/LMO ($m$ ML)/LNO sandwiched structures are shown in Fig. 3(a). In contrast to the LNO film, the LMO film exhibits a clear XMCD signal due to its ferromagnetic nature [17, 19], which likely originates from excess oxygen and/or cation vacancies or compressive strain effects from lattice mismatch with the STO substrate [33]. At first glance, the XMCD spectra of the trilayer structures appear similar to that of the $Mn^{3+}$ ions of the LMO film. However, upon closer inspection, the Mn XMCD signals of the trilayer structures with $m$ = 6 and 12 ML have additional $Mn^{4+}$ contributions, which correspond to the shoulder features centered at around 644 eV [34]. In addition, these spectral shapes are very similar to those of the hole-doped manganite $La_{1-x}Sr_xMnO_3$ [35, 36], indicating that not only $Mn^{3+}$ but also $Mn^{4+}$ contributed to the ferromagnetism of both the 6 and 12 ML LMO layers. Meanwhile, for the 2-ML trilayer structure, the spectral shape resembles that of $La_2Ni^{2+}Mn^{4+}O_6$ [21], suggesting that its magnetization results from the valence states approaching $Mn^{4+}$.

The average effective spin moments per Mn ion [17, 30, 31] determined for the LNO/LMO ($m$ ML)/LNO sandwiched structures are plotted as a function of $m$ in Fig. 3(b), together with those of the LMO film. This figure shows that the magnetic moments of the Mn ions in the trilayer structures with $m$ = 6 ML (2.7 $\mu_B$/Mn) and 12 ML (2.4 $\mu_B$/Mn) are larger than that of the LMO film (2.0 $\mu_B$/Mn). These increases are significant because if Mn spins are fully polarized ferromagnetically, the spin moments of the trilayers with $m$ = 6 and 12 ML should be smaller than that of LMO due to their spin configurations (4 $\mu_B$ for the $Mn^{3+}$ ion with $3d^4$ high spin (HS) to 3 $\mu_B$ for $Mn^{4+}$ ion ($3d^3$ HS)).

Since transferred charges are distributed across the 3–4 ML LMO next to the heterointerface [24], the increased magnetic moment in the trilayer structures suggests that the ferromagnetism of Mn ions in the interfacial region is more stable due to charge distribution (effective hole doping) than that of Mn ions in the inner region, away from the interface, which do not experience the charge transfer. The magnetic phase diagram of the hole-doped $LaMnO_3$ [37, 38, 39] suggests that this stabilized ferromagnetism in the interface region of the LMO layer can be explained by the increased ferromagnetic double exchange (DE) interactions between $Mn^{3+}$ and $Mn^{4+}$ ions due to holes distributing from the interface. This doped-hole DE phenomenon is further supported by suppressed ferromagnetism in the trilayer structure with $m$

= 2 ML, as shown in Fig. 3(b). Its magnetization results from the valence state of almost $Mn^{4+}$, as evidenced by its spectral shape [21], and the magnetic moment of Mn ions is significantly lower than that of Mn ions in the LMO film. In the trilayer structure with $m$ = 2 ML, the holes transferred from the top and bottom LNO layers are accommodated in the 2 ML LMO sandwiched layers, and consequently, the hole doping level becomes too high to stabilize the ferromagnetic DE interactions, as observed for the overdoped region of hole-doped LMO [39].

We now turn to the origins of the interfacial ferromagnetism between the LNO and LMO layers. Recent studies on LNO/LMO superlattices have also reported this interfacial ferromagnetism, which was attributed to the ferromagnetic SE interaction of $Ni^{2+}(d^8)$–O–$Mn^{4+}(d^3\ HS)$ bonds, which obey the KG rules [17, 29], analogously to the double-perovskite oxide $La_2Ni^{2+}Mn^{4+}O_6$. In general, in (111)-oriented $(LNO)_1/(LMO)_1$ superlattices that experience the charge transfer reaction $Ni^{3+} + Mn^{3+} \rightarrow Ni^{2+} + Mn^{4+}$, the rock-salt-type ordering of $Ni^{2+}O_6$ and $Mn^{4+}O_6$ octahedra occurs, which stabilizes the ferromagnetic SE interaction of $Ni^{2+}$–O–$Mn^{4+}$ bonds in all directions. On the other hand, the (001)-oriented LNO/LMO heterostructures contain not only out-of-plane $Ni^{2+}$–O–$Mn^{4+}$ bonds across the interface but also in-plane $Ni^{2+}$–O–$Ni^{2+}$ and $Mn^{4+}$–O–$Mn^{4+}$ bonds. Within the framework of the KG rules, the spins of the Mn and Ni ions in the out-of-plane $Ni^{2+}(d^8)$–O–$Mn^{4+}(d^3\ HS)$ bonds couple ferromagnetically to each other due to SE interactions, while those in the in-plane $Ni^{2+}(d^8)$–O–$Ni^{2+}(d^8)$ and $Mn^{4+}(d^3\ HS)$–O–$Mn^{4+}(d^3\ HS)$ bonds align antiferromagnetically. Consequently, as schematically illustrated in Fig. 4(a), the macroscopic magnetization of Mn and Ni ions is expected to be absent at the (001)-oriented LNO/LMO heterointerface.

In contrast, we observed distinct XMCD signals for both Ni and Mn ions at the chemically sharp (001)-oriented heterointerfaces between LNO and LMO. Our experimental findings can be summarized as follows: (i) magnetic moments reside in both Ni and Mn ions;

(ii) the magnetization of Ni ions is induced only in $Ni^{2+}$ ions in the 1 ML LNO adjacent to the interface, and their magnetic moment is much smaller than the full spin moment estimated from the electron configuration of $Ni^{2+}$ states; (iii) the magnetization degree of Mn ions in the interfacial region characterized by charge redistribution is higher than that in the inner region of LMO; and (iv) the induced and increased magnetizations of Ni and Mn ions, respectively, are coupled ferromagnetically. Based on these observations, we proposed a possible mechanism for the emergence of the ferromagnetism at the (001)-oriented LNO/LMO heterointerface, as schematically illustrated in Fig. 4(b). As previously described, the ferromagnetic DE interactions between $Mn^{3+}(d^4\,HS)$ and $Mn^{4+}(d^3\,HS)$ ions stabilize the ferromagnetic coupling in the interfacial region of the LMO layer. As for the LNO layer, according to the KG rules, the spins of $Ni^{2+}$ ions within the LNO (001) layer are expected to couple with each other in antiparallel owing to the strong antiferromagnetic SE interaction of $Ni^{2+}(d^8)$–O–$Ni^{2+}(d^8)$ bonds, as shown by the outline black arrows in Fig. 4(b). Indeed, the antiferromagnetic SE interactions in nickelates are very strong, as indicated by the high Néel temperatures of $La_2NiO_4$ ($T_N \sim 650$ K) [40] and NiO ($T_N \sim 523$ K) [41]. Meanwhile, along the out-of-plane direction, ferromagnetic SE interactions obeying the KG rules should exist between Ni and Mn ions bonded via oxygen across the heterointerface, mostly $Ni^{2+}(d^8)$–O–$Mn^{4+}(d^3\,HS)$ bonds [22, 28, 42] and $Ni^{2+}(d^8)$–O–$Mn^{3+}(d^4\,HS)$ bonds. As a result, the delicate balance between the out-of-plane and in-plane exchange interactions may cause spin canting of Ni ions within the 1 ML LNO at the interface, as shown by the hatched green arrows in Fig. 4(b). In other words, the observed weak ferromagnetism of $Ni^{2+}$ ions is likely caused by the canted ferromagnetism of $Ni^{2+}$ ions.

These results suggest that the stabilization of the ferromagnetism in LMO layers due to the interfacial charge redistribution and simultaneous ferromagnetic coupling between Ni and Mn spins are the key factors inducing the interfacial magnetism of the (001)-oriented LNO/LMO heterointerface. The model in Fig. 4(b) suggests that the appropriate LMO thickness that stabilizes the ferromagnetism is also necessary to magnetize the Ni ions in the LNO layer of the (001)-oriented LNO/LMO heterostructure. Thus, the LNO/LMO (2 ML)/LNO sandwiched structure, in which the magnetic moment of Mn ions is significantly suppressed, exhibits a negligible XMCD intensity at the Ni absorption edge (not shown). These results are consistent with those of a previous report [29], indicating that Ni ions do not magnetize when the ferromagnetic interactions in Mn ions are suppressed. In order to more comprehensively understand the origins of interfacial magnetism due to the charge transfer between the constituent transition-metal oxides, further systematic investigation is required. In particular, the effects of different constituent transition metals on the charge spreading and the resultant interfacial magnetism are especially important to examine.

## IV. CONCLUSION

We investigated the effect of the asymmetric charge redistribution at the LNO/LMO heterointerface on its ferromagnetism using element-selective XAS and XMCD techniques. Systematic thickness-dependent measurements revealed that the spatial distributions of the transferred charges strongly influence the emergence of ferromagnetism at the LNO/LMO heterointerface. In particular, weak magnetization is observed only for the $Ni^{2+}$ ions in the 1 ML LNO adjacent to the heterointerface, owing to the confinement of transferred electrons. Meanwhile, the ferromagnetism of the LMO layer is higher within 3–4 ML near the interface than those in the LMO inner region, further from the interface. Furthermore, the magnetic moments of both ions are ferromagnetically coupled through the interface. The characteristic length scale of the emergent (increased) interfacial ferromagnetism in the LNO (LMO) layers is in good agreement with the charge distribution across the interface, indicating the close relationship between the charge redistribution due to the interfacial charge transfer and the ferromagnetism of the LNO/LMO interface. Based on these observations, we proposed a possible model to describe the interfacial magnetism of the LNO/LMO heterointerface. According to this model, the weak magnetization of $Ni^{2+}$ ions in the LNO layer adjacent to the interface probably originates from the spin canting caused by the competition between the in-plane antiferromagnetic and out-of-plane ferromagnetic interactions. To better understand the interfacial ferromagnetism of these oxide heterostructures, not only the charge transfer but also its spatial distributions of the transferred charges must be considered.

*Acknowledgments*

This work was supported by a Grant-in-Aid for Scientific Research (16H07442, 18K14130, 16H02115, B25287095, 16KK0107, 16K05033, 15H02109, and S22224005) from the Japan Society for the Promotion of Science (JSPS), a Grant-in-Aid for quantum-beam research from the Institute of Materials Structure Science (IMSS), KEK, and the MEXT Element Strategy Initiative to Form Core Research Center. MK acknowledges financial support from JSPS for Young Scientists. The work performed at KEK-PF was approved by the Program Advisory Committee (proposals 2014T002 and 2015S2-005) at the IMSS, KEK.

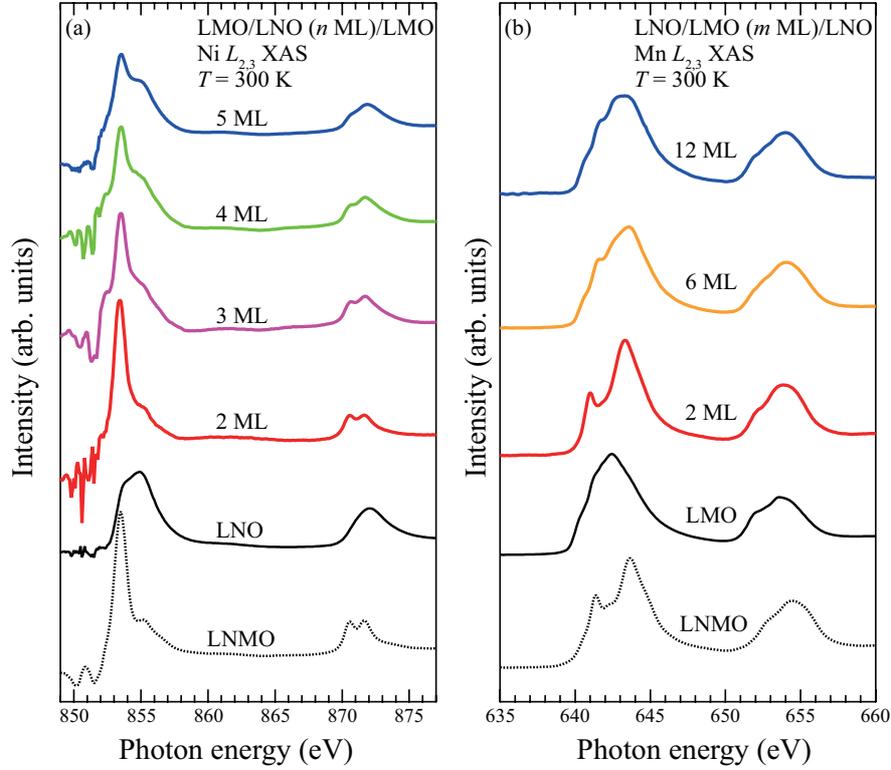

FIG. 1. (color online) (a) Ni-$L_{2,3}$ XAS spectra of the LMO/LNO ($n$ ML)/LMO trilayer structures with the sandwiched LNO layers of varying thicknesses ($n$ =2, 3, 4, and 5). The Ni-$L_{2,3}$ XAS spectra of the 20 ML LNO film and an LNMO film are also shown as references for the $Ni^{3+}$ and the $Ni^{2+}$ states, respectively. (b) Mn-$L_{2,3}$ XAS spectra of the LNO/LMO ($m$ ML)/LNO trilayer structures ($m$ =2, 6, and 12). The spectra of the 20 ML LMO film and an LNMO film are shown as references for the $Mn^{3+}$ and the $Mn^{4+}$ states, respectively. All XAS measurements were carried out at 300 K.

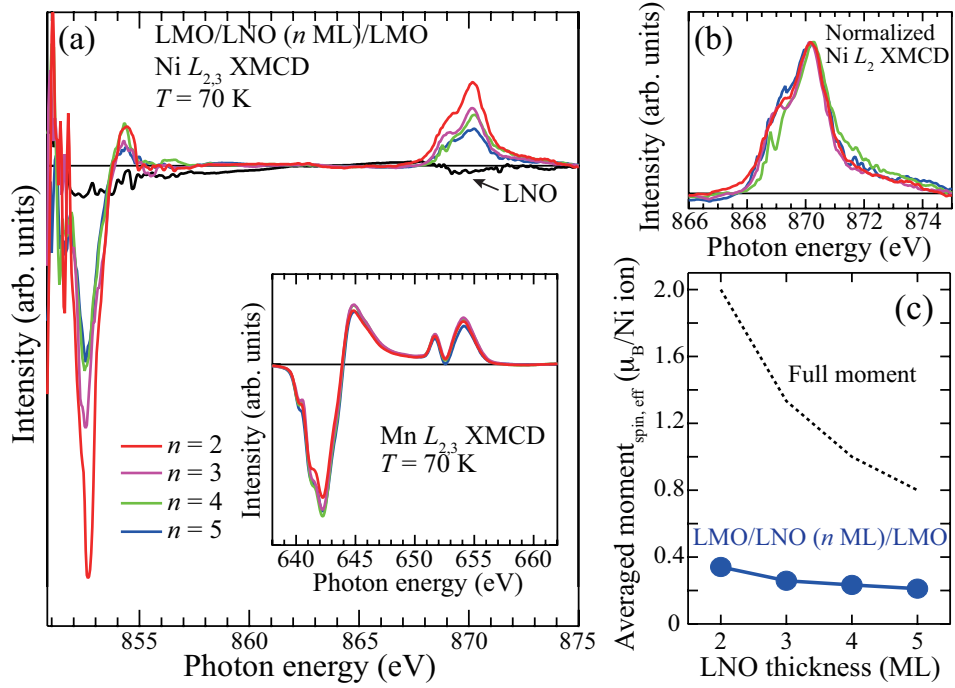

FIG. 2. (color online) (a) Ni-$L_{2,3}$ XMCD spectra of the LMO/LNO ($n$ ML)/LMO trilayer structures ($n$ =2, 3, 4, and 5) measured under the magnetic field of 1 T at 70 K. The spectrum of a 20-ML LNO film is shown for reference. The corresponding Mn-$L_{2,3}$ XMCD spectra of the trilayers are shown in the inset. (b) XMCD spectra obtained around the Ni-$L_2$ edge using an expanded energy scale, wherein the Ni-$L_2$ XMCD signals are normalized by the spectral intensity. (c) Average effective spin moments per Ni ion obtained for the LMO/LNO ($n$ ML)/LMO trilayers (closed circles with a line), together with the hypothetical average spin moment in the case that Ni$^{2+}$ ions only in the 1 ML LNO at the interface exhibit the full moment, whereas the Ni$^{3+}$ ions are not magnetized in the inner layers (dashed line).

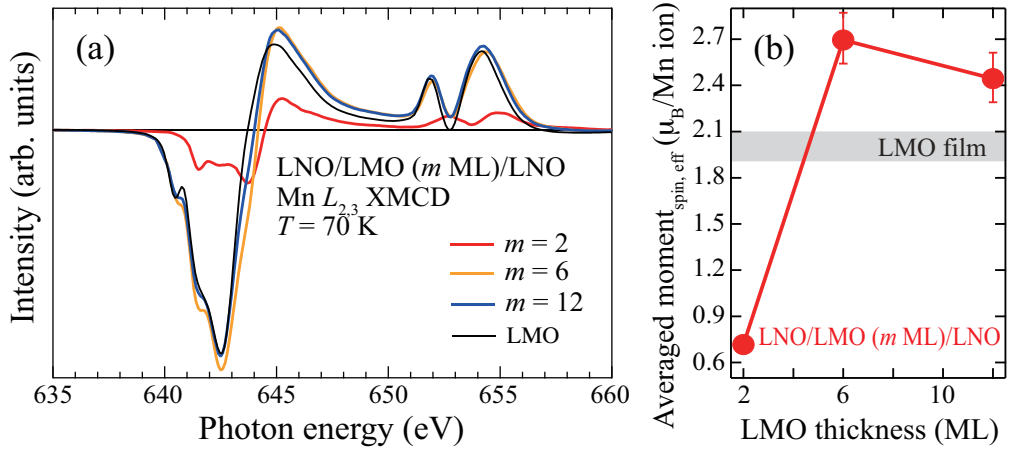

FIG. 3. (color online)   (a) Mn-$L_{2,3}$ XMCD spectra of the LNO/LMO ($m$ ML)/LNO trilayer structures ($m$ = 2, 6, and 12) obtained under the magnetic field of 1 T at 70 K.   The spectrum of a 20-ML LMO film is shown for reference.   (b) Average effective spin moments per Mn ion determined for the sandwiched LMO layers in the LNO/LMO ($m$ ML)/LNO trilayer structures. The spin moment of a 20-ML LMO film is shown for comparison.

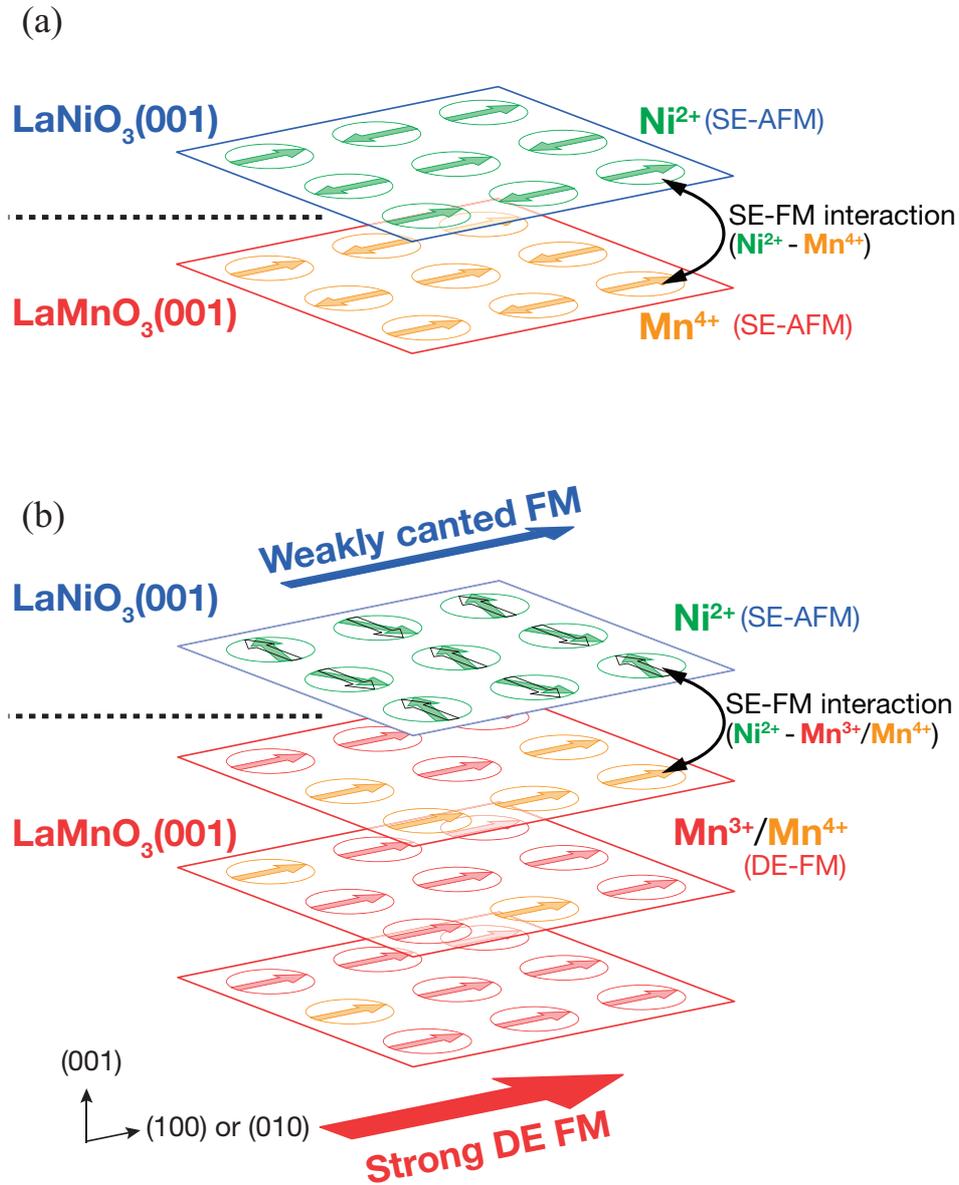

FIG. 4. (color online) Schematic illustrations of possible interfacial magnetic structures (a) without and (b) with charge redistributions. The outline black arrows in panel (b) denote the magnetic structure, in which no magnetic exchange interactions occur between Ni and Mn ions. Here, SE denotes superexchange interaction, DE means double exchange interaction, AFM is antiferromagnetism, and FM means ferromagnetism.